\documentclass[preprint,prb,amsmath,amssymb,superscriptaddress]{revtex4-1}
\usepackage{hyperref}
\usepackage{graphics}
\usepackage{enumerate}

\begin{document}


\section*{Single Atom Qubits: Acceptors}
\noindent J. Salfi\\
\noindent Department of Electrical and Computer Engineering, The University of British Columbia. 2332 Main Mall, Vancouver, BC. V6T1Z4, Canada.
\subsection*{Status}
\begin{figure*}
\includegraphics{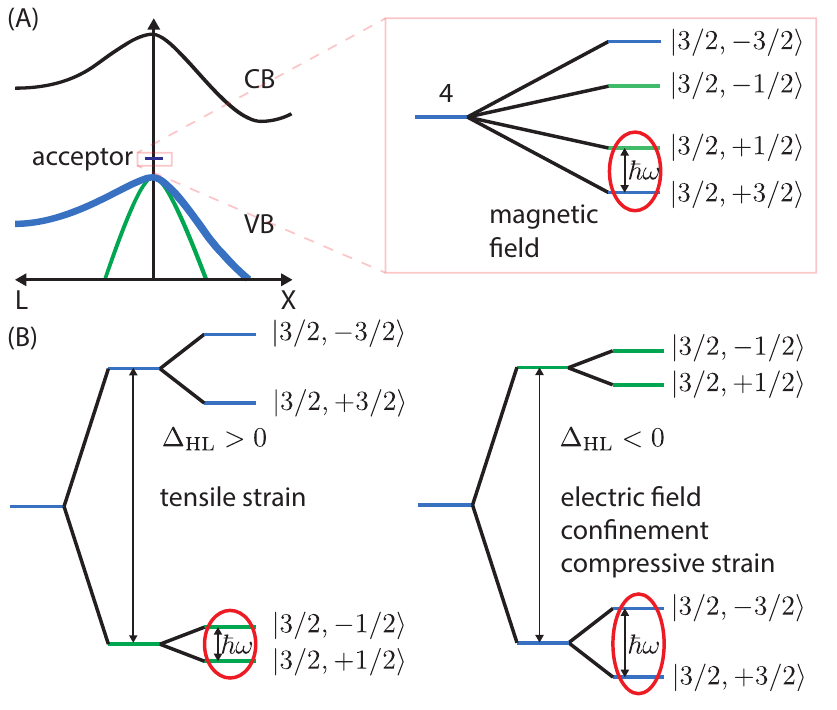}
		\caption
		{(A) Acceptor bound state above the valence band edge. Levels labelled $\left|J,m_J\right\rangle$, where $J$ is the total angular momentum and $m_J$ is the projected total angular momentum are split into four in a magnetic field, in a Si crystal. The lowest energy states form a two-level system with a Larmor frequency $\hbar\omega$. (B) Time-reversal symmetric two-level quantum systems are induced when a gap $\Delta_{\rm HL}$ is induced. $\left|3/2,\pm 1/2\right\rangle$ is obtained under biaxial tensile strain, and is obtained under compressive strain, confinement, or electric fields, which couple to electric and elastic fields\cite{Ruskov2013, Salfi2016b, AbadilloUriel2018,Golding2003}.}\label{fig1}
\end{figure*}

\noindent Acceptor dopant atoms have recently been identified as compelling candidates for spin-based quantum technologies. Interest in acceptors ultimately derives from the properties of their acceptor-bound holes (Fig.~\ref{fig1}A), where spin-orbit coupling quantizes total angular momentum $J=3/2$ rather than spin. Under applied magnetic, electric, and elastic fields, different two-level systems can be defined (Fig.~\ref{fig1}B) amenable to two-qubit logic over long distances \cite{Ruskov2013, Salfi2016b, AbadilloUriel2018,Golding2003} and fast single-qubit logic using electric fields \cite{Salfi2016b,AbadilloUriel2018}. These properties are important to improve the scalability of spin-based technologies, and here derive from spin-orbit coupling, which is weak for electrons in silicon. Two-qubit operations are predicted to be possible either indirectly, using microwave phonons\cite{Ruskov2013} or microwave photons\cite{Salfi2016b} in cavity quantum electrodynamics (QED), or via elastic dipole-dipole \cite{Golding2003} or electric dipole-dipole \cite{Salfi2016b} interactions, even while suppressing decoherence from electric field noise \cite{Salfi2016b}. Phonon coupling could enable transducers from microwave to optical photons for quantum networks \cite{Ruskov2013}. 

Experimental investigation of acceptor-bound holes is underway with B:Si acceptors and is confirming their potential for quantum technologies. The first single atom transistor was demonstrated in an industrially fabricated device\cite{Vanderheijden2014}, exhibiting the $J=3/2$ Zeeman energy spectrum (Fig.~\ref{fig1}A). Readout by spin-to-charge conversion was also demonstrated on an industrially fabricated two-atom device by gate-based reflectometry\cite{Vanderheijden2018}. Recent materials advances are also paving the way. Isotope purification, which removes random strains in the host, yields narrow linewidths in ensemble continuous wave spin resonance\cite{Stegner2010}. Recently, pulsed spin resonance has been performed for B acceptors in $^{28}$Si yielding ultra-long 10 ms spin coherence times $T_2$ in a moderate static strain\cite{Kobayashi2018}, approaching the best results for electron spins. This is highly non-trivial because spin and orbital are mixed in $J=3/2$ systems, but is a key enabling ingredient making acceptors attractive for quantum technologies.

\subsection*{Current and Future Challenges}

\begin{figure*}
\includegraphics{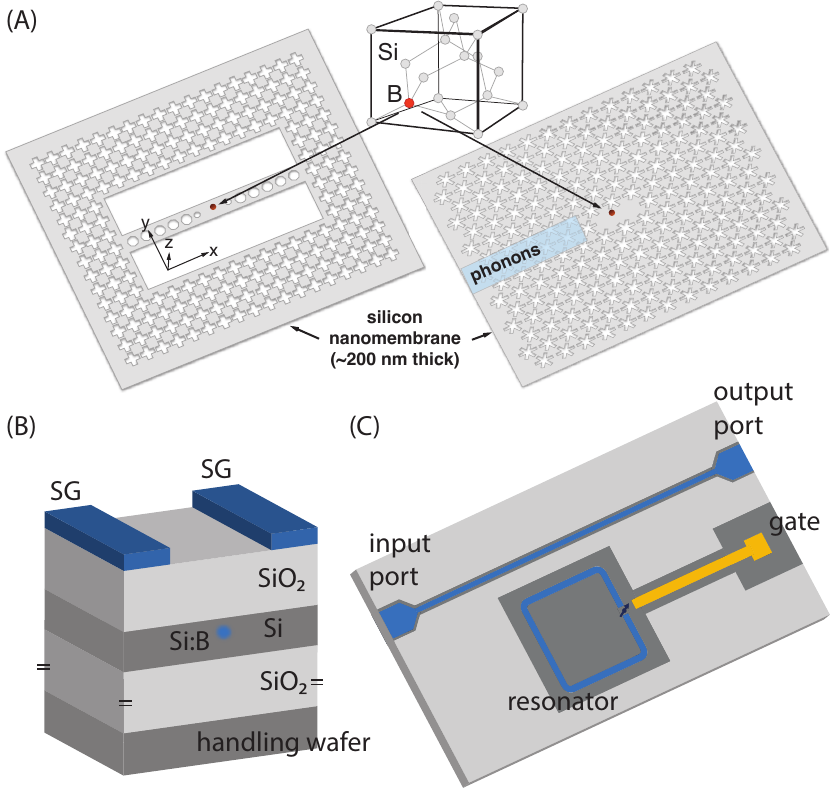}
		\caption
		{(A) quantum system of a Si acceptor and a one-dimensional (left) and two-dimensional (right) nanomechanical cavity. An on-chip phonon waveguide could be used to couple to the hybrid system. The elastic field in the nanomechanical resonator couples to the acceptor\cite{Ruskov2013} (B) Si acceptor in a strained silicon-on-insulator structure with side gates SG (left) (C) Hybrid quantum system of a Si acceptor and a nanowire ring-based high impedance superconducting resonator. Microwave photons in the resonator couple to the acceptor  [2].}\label{fig2}
\end{figure*}

The next breakthrough required to establish the suitability of acceptor-based qubits for quantum computing is to couple acceptor qubits in scalable arrays. Two-dopant atom coupling for acceptors has been demonstrated using the exchange mechanism\cite{Vanderheijden2018,Salfi2016a}. While useful for two-qubit gates, exchange is short-ranged making it difficult to fabricate two-dimensional qubit arrays and the desired measurement and control devices needed for quantum error correction. Coupling via electric or elastic fields is therefore more attractive. Devices allowing applied strain fields, electric fields, and interaction with interfaces/confinement (Fig.~\ref{fig1}B) is predicted to enable control over electric and elastic couplings needed for the long-ranged coupling, via control of energy gaps\cite{Ruskov2013, Salfi2016b, AbadilloUriel2018,Golding2003} and Rashba-like interactions\cite{Salfi2016b,AbadilloUriel2018}. Indirect QED-based schemes where interactions are mediated by phonons or photons will require nanomechanical (Fig.~\ref{fig2}A) and superconducting (Fig.~\ref{fig2}B,C) cavity design to obtain the desired spin-to-photon or spin-to-phonon coupling, respectively. It also requires integration of acceptor atoms into these cavities, but could allow qubit readout with essentially no overhead.

Another breakthrough will be realizing gate fidelities above typically 99~\% or higher required for large-scale quantum computers that are tolerant to errors. The long $T_2$ of acceptors is advantageous for this, because infidelity is bounded by $\tau/T_2$, where $\tau$ is the gate time, which is expected to be in the $\sim 10$~ns range\cite{Salfi2016b,AbadilloUriel2018}. One of the challenges will be to maintain the long $T_2$ times of acceptors when long-range couplings are used. The use of optimal working points (``sweet spots'') where $T_2$ is optimized but electric couplings are active have been identified in theoretical work on acceptors, and should play an important role\cite{Salfi2016b}. 

Another breakthrough would be optically interconnecting physically separated systems in quantum networks using microwave-to-optical quantum transducers. To accomplish this with acceptor atoms\cite{Ruskov2013}, optical structures are needed where the optical modes are sensitive to small mechanical deformations that couple via the spin-phonon interaction. 

\subsection*{Advances in Science and Technology to Meet Challenges}
The above challenges can be met by building acceptor-based qubit systems in a way that leverages the existing process technology and yields devices with the desired characteristics. Like for donors and quantum dots, the acceptor platform is backed by the materials and process know-how from the microelectronics industry. Within this context, acceptors are poised to take advantage of recent materials development aimed at extending Moore's law. Indeed, TiN gate materials, which in recent years have been adopted as the gate material in ultra-scaled transistors, have recently been shown to be suitable for building superconducting resonators with high quality factors and high characteristic impedances\cite{Shearrow2018} desirable for QED via electrically mediated interactions with spin qubits. Strain control, which enables the enhancement of acceptor $T_2$ to state-of-the-art values, also features in state-of-the-art microelectronic devices. Generally, there are two ways to achieve this, either to use strained silicon-on-insulator (Fig.~\ref{fig2}B) or to use the gate material itself, such as TiN, to controllably strain the lattice. There is already an advanced research and industrial fabrication infrastructure in place for silicon-based nanomechanical and nanophotonic structures to build microwave-to-optical quantum transducers for quantum networks with high-quality mechanical and photonic components. Existing silicon-based technologies form a solid basis to investigate scalable quantum information technologies with acceptor qubits. 

\subsection*{Concluding Remarks}
Acceptor-based hole spins offer compatibility with established silicon fabrication techniques together with recently demonstrated long spin lifetimes in $^{28}$Si. Their potential for addressable electric spin manipulation and long-distance coupling via electric or elastic fields and makes them compelling new candidates for scalable quantum computers and networks.

\subsection*{Acknowledgements}
J. Salfi acknowledges J. T. Muhonen for helpful comments and M. Khalifa for the schematic of the superconducting microwave resonator. Funding support for this work was provided by the National Science and Engineering Research Council of Canada and the Canadian Foundation for Innovation. 


\end{document}